\documentclass[12pt]{article}
\usepackage{epsfig}

\def\beq{\begin{equation}}
\def\eeq#1{\label{#1}\end{equation}}
\def\eeqn{\end{equation}}

\def\beqa{\begin{eqnarray}}
\def\eeqa#1{\label{#1}\end{eqnarray}}
\def\eeqan{\end{eqnarray}}

\let\bar=\overbar

\def\Dslash{\not{\hbox{\kern-4pt $D$}}}
\def\dslash{\not{\hbox{\kern-2pt $\del$}}}

\def\msb{{\bar{\ssstyle M \kern -1pt S}}}

\def\Title#1{\begin{center} {\Large {\bf #1} } \end{center}}

\usepackage{relsize}
\def\babar{\mbox{\slshape B\kern-0.1em{\smaller A}\kern-0.1em
    B\kern-0.1em{\smaller A\kern-0.2em R}}}
\usepackage{xspace}

\def\CP    {\ensuremath{C\!P}\xspace}
\def\CPT   {\ensuremath{C\!PT}\xspace}

\def\pim   {\ensuremath{\pi^-}\xspace}

\def\KS    {\ensuremath{K^0_{\scriptscriptstyle S}}\xspace} 
\def\KL    {\ensuremath{K^0_{\scriptscriptstyle L}}\xspace} 
\def\Kp    {\ensuremath{K^+}\xspace}

\def\Kz      {\ensuremath{K^0}\xspace}
\def\Kbar    {\kern 0.2em\overline{\kern -0.2em K}{}\xspace}

\def\Kzb     {\ensuremath{\Kbar^0}\xspace}
\def\KzKzb   {\ensuremath{\Kz \kern -0.16em \Kzb}\xspace}

\newcommand{\gev}{\ensuremath{\mathrm{\,Ge\kern -0.1em V}}\xspace}
\newcommand{\mev}{\ensuremath{\mathrm{\,Me\kern -0.1em V}}\xspace}
\newcommand{\gevc}{\ensuremath{{\mathrm{\,Ge\kern -0.1em V\!/}c}}\xspace}
\newcommand{\mevc}{\ensuremath{{\mathrm{\,Me\kern -0.1em V\!/}c}}\xspace}
\newcommand{\gevcc}{\ensuremath{{\mathrm{\,Ge\kern -0.1em V\!/}c^2}}\xspace}
\newcommand{\mevcc}{\ensuremath{{\mathrm{\,Me\kern -0.1em V\!/}c^2}}\xspace}

\def\ccbar   {\ensuremath{c\overline c}\xspace}

\def\B       {\ensuremath{B}\xspace}
\def\Bbar    {\kern 0.18em\overline{\kern -0.18em B}{}\xspace}

\def\Bz      {\ensuremath{B^0}\xspace}
\def\Bzb     {\ensuremath{\Bbar^0}\xspace}
\def\BzBzb   {\ensuremath{\Bz {\kern -0.16em \Bzb}}\xspace}
\def\Bu      {\ensuremath{B^+}\xspace}
\def\Bub     {\ensuremath{B^-}\xspace}

\def\BpBm    {\ensuremath{\Bu {\kern -0.16em \Bub}}\xspace}

\def\jpsi     {\ensuremath{{J\mskip -3mu/\mskip -2mu\psi\mskip 2mu}}\xspace}

\mathchardef\Upsilon="7107
\def\Y#1S{\ensuremath{\Upsilon{(#1S)}}\xspace}

\def\Bcpp  {\ensuremath{\B_{+}}\xspace}
\def\Bcpm  {\ensuremath{\B_{-}}\xspace}

\begin{document}

\Title{Observation of Time-reversal Violation at \babar}

\bigskip\bigskip


\begin{raggedright}  

{\it Ray F. Cowan\index{Cowan, Ray F.}\footnote{Speaker on behalf of the \babar\ Collaboration.}\\
Laboratory for Nuclear Science\\
M.I.T.\\
Cambridge, Masschusetts 02139 USA\par}
\bigskip\bigskip
Proceedings of CKM 2012, the 7th International Workshop on the CKM Unitarity Triangle, University of Cincinnati, USA, 28~September -- 2~October 2012 
\end{raggedright}

\section{Time-reversal violation}

In stable systems, violation 
of {\it Umkehr der Bewegungsrichtung}~\cite{Wigner:1957ep,Luders:1952ab},
or time-reversal ($T$), symmetry
is indicated by a non-zero value of $T$-odd observables such as
the neutron or electron electric dipole moments 
($d_n < 2.9\times 10^{-26}$~e-cm, $d_e < 10.5 \times 10^{-28}$~e-cm, 
at 90\% CL).\cite{Beringer:1900zz}
$T$~violation would also be indicated by different probabilities 
for $a\to b$ at a given time than for $b\to a$, such as $\nu_e \to \nu_{\mu}$
vs. $\nu_{\mu} \to \nu_e$ at a muon storage ring.

In unstable systems, $T$ violation can be explored by 
study of a process under the transformation 
$t \to -t$ combined with exchange of
$|in\rangle$ and $|out\rangle$ states, which can be experimentally
challenging to achieve. Comparing the rates of $\Bz\to\Kp\pim$ and 
$\Kp\pim\to\Bz$ is not feasible due to the need to prepare the initial
state and to disentangle weak from strong effects.  Comparing mixing
rates $\Bz\to\Bzb$ and $\Bzb\to\Bz$ does not distinguish \CP from $T$
violation and assumes \CPT 
non-invariance~\cite{Kabir:1970ts,Angelopoulos:1998dv}.
Searches in interference
between mixing and decay ($\Bz\to\Bzb$, $\Bzb\to f_{CP}$ vs. 
$\Bz\to f_{CP}$) does not exchange $|in\rangle$ and $|out\rangle$ states
or $t\to -t$ and assumes \CPT non-invariance and $\Delta\Gamma=0$.

\section{Experimental procedure}
We use the entangled quantum state $|i\rangle$
of two \B mesons from an \Y4S
decay:
\begin{eqnarray}
|i\rangle & = & 1/\sqrt{2}\left[\Bz(t_1)\Bzb(t_2) - 
\Bzb(t_1)\Bz(t_2)\right]\nonumber\\
 & = & 1/\sqrt{2}\left[B_+(t_1)B_-(t_2) - B_-(t_1)B_+(t_2)\right]\nonumber
\end{eqnarray}
where $\Bcpp$ and $\Bcpm$ are mutually orthogonal states decaying
to the \CP eigenstates $\jpsi\KL$ and $\jpsi\KS$ ($\KS\to\pi\pi$).
with $\CP=+$ and $\CP=-$, respectively.
We define reference transitions and their $T$-transformed counterparts
as shown in Table~\ref{tab:one}.  We use the notation $(X,Y)$ to denote
the final states of both $B$ mesons which decay from an entangled state
where $\B\to X$ is the first decay and $\B\to Y$ is the second decay such
that the decay time difference is positive by construction: 
$\Delta t = t_Y - t_X > 0$.  As an example, the final state $(l^+,\jpsi\KS)$
represents 
a $\Bz\to l^+X$ decay followed in time by a $\Bzb\to\Bcpm$ decay. We know
the second \B starts in a \Bzb flavor state from the orthogonality of the
initial state and its \CP eigenvalue at decay is indicated by its decay mode.
A difference
between this rate and that of 
the time-reversed transition $\Bcpm\to\Bzb$, with final state 
$(\jpsi\KL,l^-)$, is an indication of $T$~violation.
A total of four $T$-reversed transitions can be studied 
(see Table~\ref{tab:one}).  Four other final-state combinations can
be compared to study \CP violation and another four for \CPT violation,
all independent.

\begin{table}[t]
\begin{center}
\begin{tabular}{cc}
\hline
Reference transition $(X,Y)$& $T$-transformed transition $(X,Y)$\\
\hline
$\Bz\to\Bcpp$ $(l^-,\jpsi\KL)$ & $\Bcpp\to\Bz$ $(\jpsi\KS,l^+)$\\
$\Bz\to\Bcpm$ $(l^-,\jpsi\KS)$ & $\Bcpm\to\Bz$ $(\jpsi\KL,l^+)$\\
$\Bzb\to\Bcpp$ $(l^+,\jpsi\KL)$ & $\Bcpp\to\Bzb$ $(\jpsi\KS,l^-)$\\
$\Bzb\to\Bcpm$ $(l^+,\jpsi\KS)$ & $\Bcpm\to\Bzb$ $(\jpsi\KL,l^-)$\\
\hline
\end{tabular}
\caption{Reference transitions and their $T$-transformed counterparts.
$\Bcpp$ and $\Bcpm$ are orthogonal states decaying
to \CP eigenstates with $\CP=+$ and $\CP=-$, respectively; 
the notation $(X,Y)$ indicates the final states of both $B$ mesons
where $X$ is the earlier decay; and
the notation $l^{+}$ ($l^{-}$) identifies the \B meson as a \Bz (\Bzb).}
\label{tab:one}
\end{center}
\end{table}

\par
The dataset and event selection are essentially
the same as that of the \babar\ \CP violation study~\cite{Aubert:2009aw}.
The decay rate $g_{\alpha,\beta}^{\pm}(\Delta t)$ is proportional to
\begin{eqnarray}
\label{eq:rate}
g_{\alpha,\beta}^{\pm}(\Delta t)\propto e^{-\Gamma_d \Delta t} \big\{ 1+ S_{\alpha,\beta}^\pm \sin(\Delta m_d\Delta t) + C_{\alpha,\beta}^\pm \cos(\Delta m_d\Delta t) \big\}.
\end{eqnarray}
There are eight distinct sets of $S_{\alpha,\beta}^\pm$, 
$C_{\alpha,\beta}^\pm$ parameters, where $\alpha$ indicates $l^{\pm}$,
$\beta$ indicates $\KS$ or $\KL$, and $\pm$ indicates whether the
flavor final state decay occurs before or after the \CP decay $\beta$.
$\Gamma_d$ is the average decay width, $\Delta m_d$ is the mass difference
of \Bz and \Bzb.  By comparison, the standard \CP study~\cite{Aubert:2009aw}
has one set of $S$, $C$ parameters and assumes $\Delta\Gamma$ is zero.
An independent flavor sample is used to determine time-resolution parameters
and wrong-sign fit fractions.  An unbinned, maximum-likelihood fit
is performed to the \Bz, \Bzb, $\ccbar\KS$, and $\jpsi\KL$ samples,
yielding the $S_{\alpha,\beta}^\pm$, $C_{\alpha,\beta}^\pm$ parameters. 
The $T$, $\CP$, and $\CPT$ violating
parameters $\Delta S^{\pm}_i$, $\Delta C^{\pm}_i$, where
$i = T$, $CP$, or $CPT$, are constructed as the differences
in $S_{\alpha,\beta}^\pm$ and $C_{\alpha,\beta}^\pm$ for
symmetry-transformed transitions (e.g., $\Delta S_T^+ = S_{l^-,\KL}^- - S_{l^+,\KS}^+$,
$\Delta C_{CPT}^- = C_{l^+,\KL}^+ - C_{l^+,\KS}^-$).

\section{Results and conclusion}

\begin{table}[htb]
\hbox to\hsize{\hss
\begin{tabular}{cccccc}
\hline
Param.& Final result & Param. & Final result & Param. & Final result \\
\hline
$\Delta S^+_{T}$ & $-1.37\pm0.14\pm0.06$& $\Delta S^+_{CP}$ & $-1.30\pm0.10\pm0.07$ & $\Delta S^+_{CPT}$ & $\phantom{-}0.16\pm0.20\pm0.09$ \\
$\Delta S^-_{T}$ & $\phantom{-}1.17\pm0.18\pm0.11$& $\Delta S^-_{CP}$ & $\phantom{-}1.33\pm0.12\pm0.06$ & $\Delta S^-_{CPT}$ & $-0.03\pm0.13\pm0.06$ \\
$\Delta C^+_{T}$ & $\phantom{-}0.10\pm0.16\pm0.08$& $\Delta C^+_{CP}$ & $\phantom{-}0.07\pm0.09\pm0.03$ & $\Delta C^+_{CPT}$ & $\phantom{-}0.15\pm0.17\pm0.07$ \\
$\Delta C^-_{T}$ & $\phantom{-}0.04\pm0.16\pm0.08$& $\Delta C^-_{CP}$ & $\phantom{-}0.08\pm0.10\pm0.04$ & $\Delta C^-_{CPT}$ & $\phantom{-}0.03\pm0.14\pm0.08$ \\
\hline
\end{tabular}\hss}%
\caption{Central values of parameters from the $T$, $\CP$, and $CPT$ fits.
The first uncertainty is statistical, the second is systematic.}
\label{tab:two}
\end{table}

\begin{figure}[htb]
\hbox to\hsize{\hss
\begin{tabular}{ccc}
\includegraphics[width=0.35\textwidth]{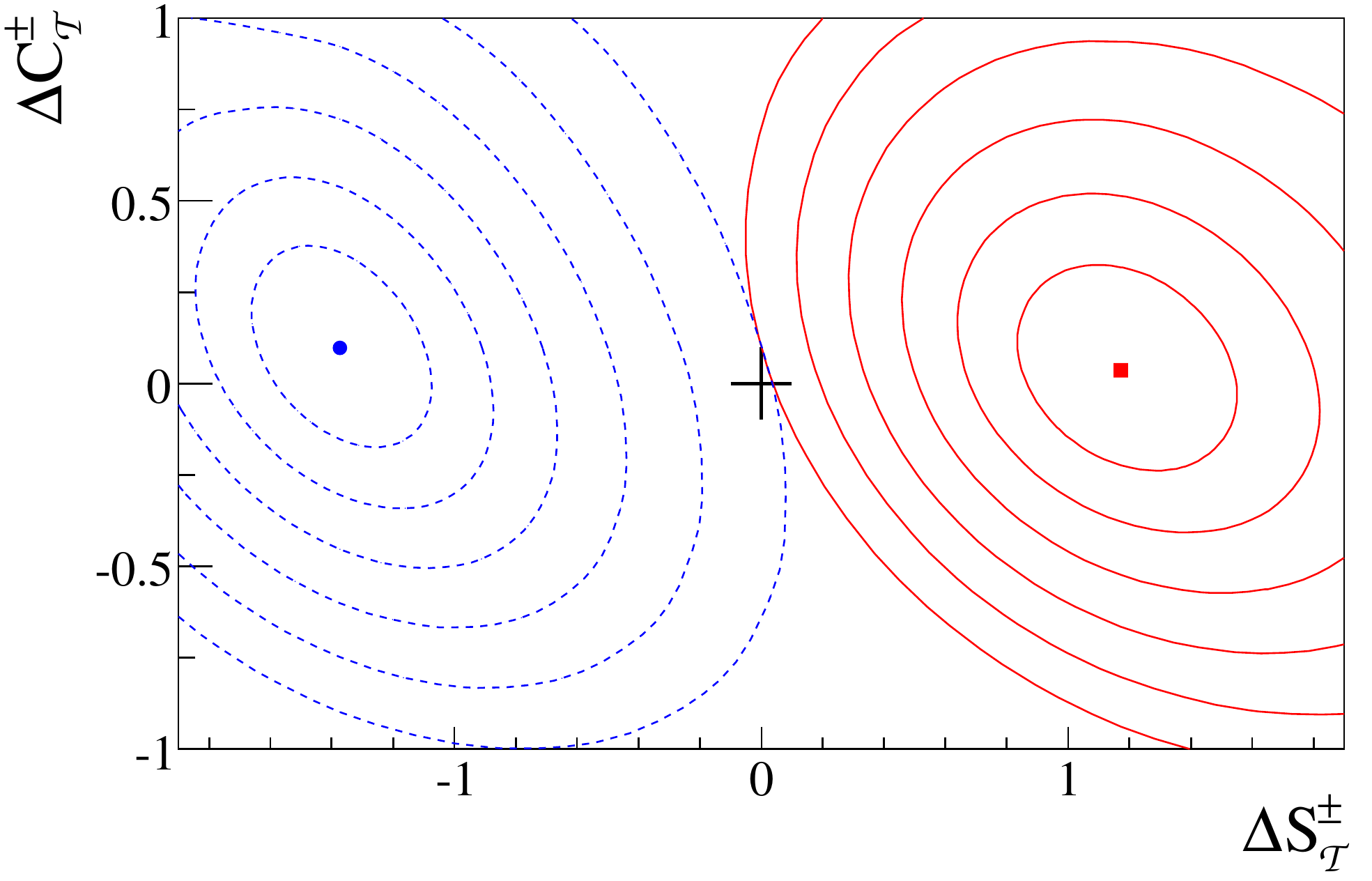} &
\includegraphics[width=0.35\textwidth]{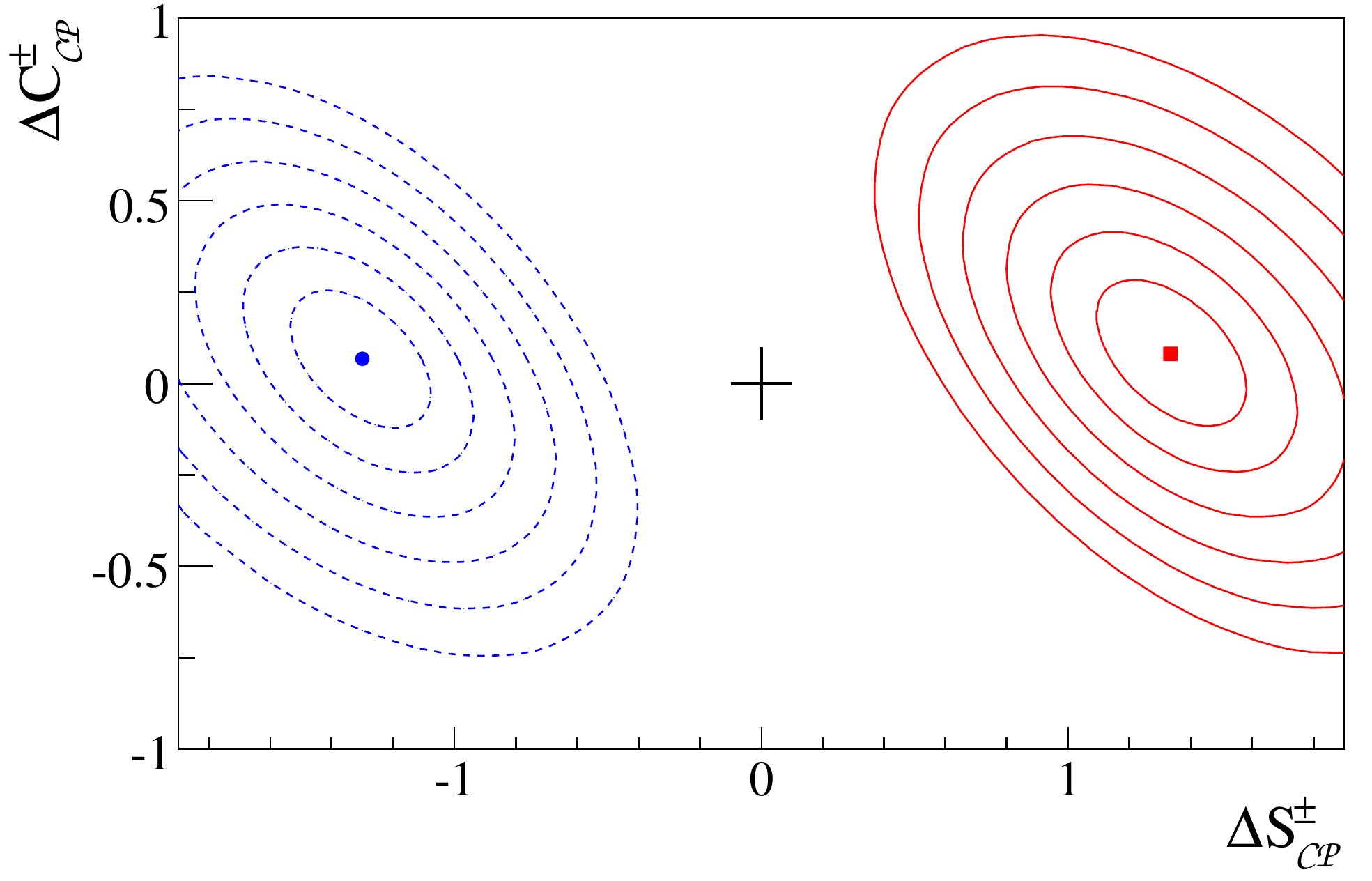} &
\includegraphics[width=0.35\textwidth]{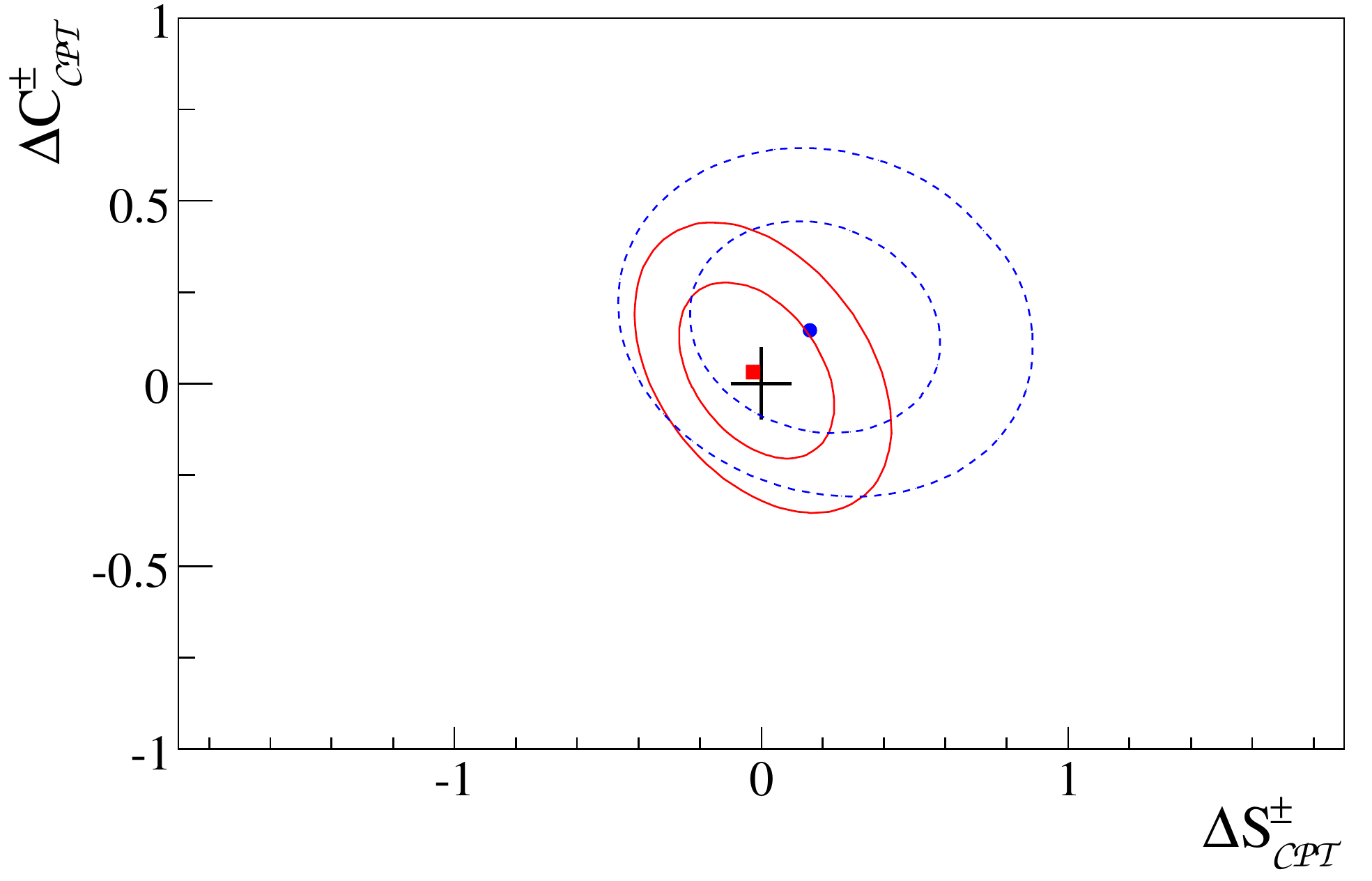}
\end{tabular}\hss}
\caption{Central values (blue points and red squares) and contours of $1-{\rm C.L.}$
at $1\sigma$ intervals for the $T$ (left), $\CP$ (middle), and $CPT$ (right)
results.  Systematic uncertainties are incorporated (the largest being 
the fit bias from simulation). Contours for 
$\Delta S^+_i$, $\Delta C^+_i$ ($i=T$, $CP$, or $CPT$) are shown as blue dashed curves.   Contours for
$\Delta S^-_i$, $\Delta C^-_i$ ($i=T$, $CP$, or $CPT$) are shown as red solid curves. The no-violation
point is shown in each plot as a plus sign ($+$).}
\label{fig:one}
\end{figure}

Results are shown in Table~\ref{tab:two}.  Their confidence level 
significances can be shown 
graphically as two-dimensional contour plots for $\Delta S^{\pm}_i$
and $\Delta C^{\pm}_i$, $i=T$, $CP$, or $CPT$ (see Figure~\ref{fig:one}).  
No assumptions about \CP or \CPT
violation or invariance are made.  
$T$ violation is observed at the $14\sigma$
level, consistent with measurements of \CP violation that assume \CPT
invariance. This is the first direct observation of $T$ violation in 
the \B system.  \CP violation is also observed at the $16\sigma$
level. No evidence for \CPT non-invariance is seen~\cite{Lees:2012kn}.


We are grateful for the
excellent luminosity and machine conditions made possible by our
PEP-II colleagues which have made this work possible and for 
the expertise and dedication of the computing organizations that
support \babar.
The collaborating institutions wish to thank the SLAC National
Accelerator Laboratory for its support and kind hospitality.
This work is supported by the
U.S. Department of Energy
and National Science Foundation, the
Natural Sciences and Engineering Research Council (Canada),
the Commissariat \`a l'Energie Atomique and
Institut National de Physique Nucl\'eaire et de Physique des Particules
(France), the
Bundesministerium f\"ur Bildung und Forschung and
Deutsche Forschungsgemeinschaft
(Germany), the
Istituto Nazionale di Fisica Nucleare (Italy),
the Foundation for Fundamental Research on Matter (The Netherlands),
the Research Council of Norway, the
Ministry of Education and Science of the Russian Federation,
Ministerio de Educaci\'on y Ciencia (Spain), and the
Science and Technology Facilities Council (United Kingdom).
Individuals have received support from
the Marie-Curie IEF program (European Union) and
the A. P. Sloan Foundation (USA), and the Binational Science
Foundation (USA-Israel).

\bibliographystyle{unsrt}
\bibliography{Cowan_BaBar_TRV_WG_IV_CKM2012}


\def\Discussion{
\setlength{\parskip}{0.3cm}\setlength{\parindent}{0.0cm}
     \bigskip\bigskip      {\Large {\bf Discussion}} \bigskip}
\def\speaker#1{{\bf #1:}\ }
\def\endDiscussion{}

\end{document}